\documentclass[conference]{IEEEtran}

\usepackage{booktabs}
\usepackage{multirow}
\usepackage{tabularx}
\ifCLASSINFOpdf
\usepackage[pdftex]{graphicx}
\graphicspath{{./}}
\DeclareGraphicsExtensions{.pdf,.jpeg,.jpg,.png, .jpg}
\else
\fi
\usepackage{amsmath}
\usepackage{algorithmic}
\usepackage{array}
\usepackage{graphicx}
\usepackage{caption}
\usepackage{subcaption}
\usepackage{listings}
\usepackage{fixltx2e}
\usepackage{stfloats}
\usepackage[hyphens]{url}
\usepackage{listings}
\begin{document}
\title{Steal Your Life Using 5 Cents:\\
 Hacking Android Smartphones with NFC Tags}

%
\author{\IEEEauthorblockN{Carlos Bermejo,
Pan Hui}
\IEEEauthorblockA{ System and Media Laboratory (SyMLab)\\
 Hong Kong University of Science and Technology\\
Email: \{cbf, panhui\}@cse.ust.hk}}

\maketitle

\begin{abstract}


Nowadays privacy in the connected world is a big user's concern. The ubiquity of mobile devices permits billions of users browse the web at anytime, anywhere. Near Field Communication (NFC) appeared as a seamlessly and simply communication protocol between devices. Commercial services such as Android Pay, and Apple Pay offer contactless payment methods that are spreading in more and more scenarios. However, we take risks while using NFC on Android devices, we can be hacked, and our privacy can be affected. In this paper we study the current vulnerabilities in the NFC-Android ecosystem. We conduct a series of experiments and we expose that with NFC  and Android devices are vulnerable to URL/URI spoofing, Bank/social network information hacking, and user's device tracking via fingerprint and geo-location.  It is important for the community to understand the problem and come up solution that can tackle these issues and inform the users about privacy awareness and risks on using these contactless services.

\end{abstract}


%
\IEEEpeerreviewmaketitle

\section{Introduction} \label{sec:introduction}

The web has become an essential part of our society and it is currently the main medium of information delivery. Billions of users browse the web on a daily basis, and there are single websites that have reached over one billion user accounts. In this environment, the ability to track users and their online habits can be very lucrative for advertising companies, yet very intrusive for the privacy of users. Many we services and Internet service providers (ISPs) aims to track the mobility and usage patterns of client hosts, the so-called device fingerprinting
(\cite{nikiforakis2013:cookieless}). The raising awareness of privacy concerns is illustrated in the cookies, location
browser pop-up notification messages. Device fingerprinting using the browser as the information channel is a main concern in the privacy field as it raises the conflict between adaptability and user privacy. Fingerprinting techniques are still in the newspapers as it appeared to be the recent suspicious method to track Uber\footnote{\url{www.uber.com}} Apple users\footnote{\url{https://www.nytimes.com/2017/04/23/technology/travis-kalanick-pushes-uber-and-himself-to-the-precipice.html?_r=0}}. If we want to have more user friendly individual characteristics user interface (UI) in the ubiquitous connected world, we might need to open some doors to developers to enable them to access this individual features. However, this open door is not always used to improve the user experience (UX) but to collect individual information, the fingerprint, where our traces in this connected world can be uniquely identify.

The forthcoming Internet of things (IoT) paradigm is stepping inside our lives, since modern smartphones, we have been surrounded for myriad of sensors to improve our lives such as sensor networks, wearable devices. NFC protocol has received special attention in many research studies and commercial systems as an efficient and simple approach to interact with the IoT ecosystem. NFC main features are low energy requirements and its limited data transmission in comparison with other wireless protocols such as Bluetooth, WiFi. 
Despite its prevalent use in current mobile networks, there are several existing or potential vulnerabilities of NFC protocols. In \cite{nelson2013:security}, the authors investigate a wide range of these weaknesses including eavesdropping, URI obfuscation, tag tampering, relay attacks, data corruption, man in the middle and worms. It also provides possible detection and mitigation mechanisms towards each of these susceptibilities. Most NFC communications do not include an encryption mechanism since it assumes that the short communication range (i.e., less than 4 cm) can guarantee the security. However, as we describe in Section~\ref{sec:nfc_vulnerabilities}, there are several studied vulnerabilities with the NFC protocol.

\begin{figure*}[t]
\begin{subfigure}[t]{0.5\columnwidth}
\includegraphics[width=\textwidth]{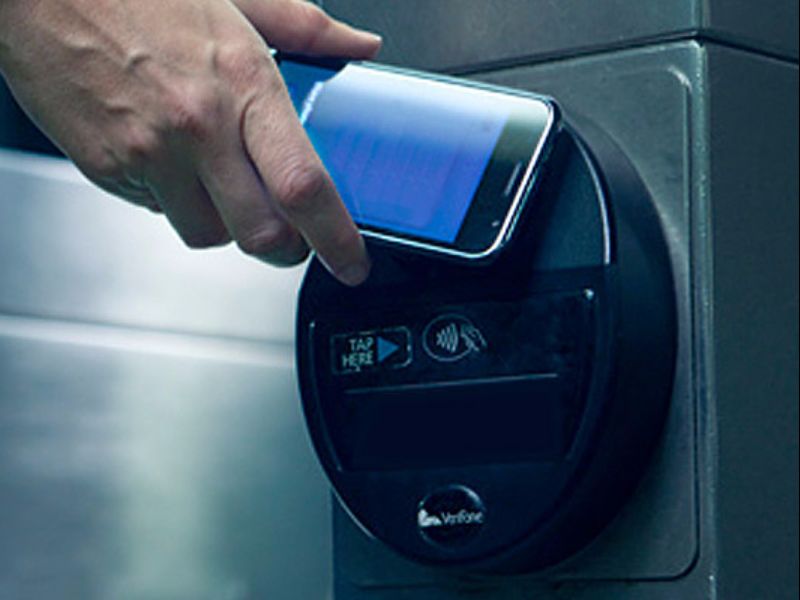}
\caption{Transport (\url{https://securityintelligence.com/is-nfc-still-a-vulnerable-technology/})}
\label{fig:nfc_transport}
\end{subfigure}
\begin{subfigure}[t]{0.5\columnwidth}
\includegraphics[width=\textwidth]{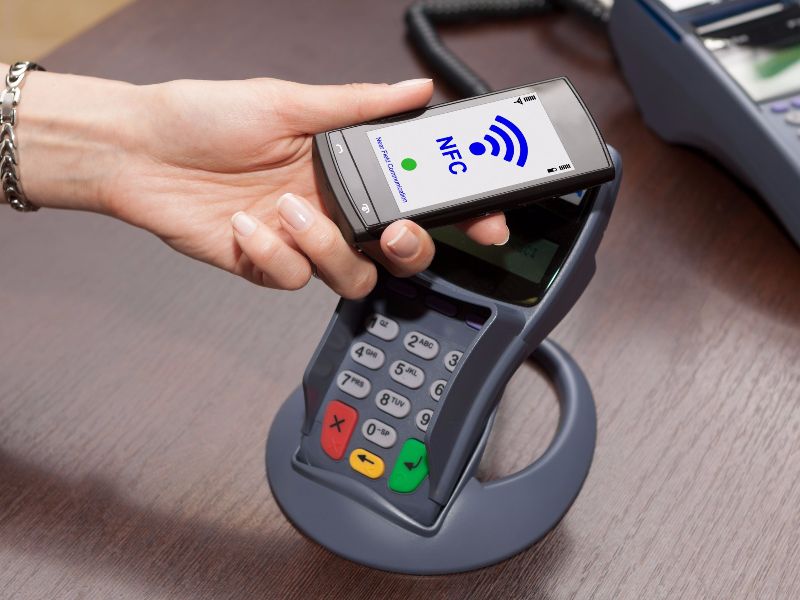}
\caption{Contactless payment~(\url{http://www.himanshutech.com/what-is-nfc/})}
\label{fig:nfc_payment}
\end{subfigure}
\begin{subfigure}[t]{0.5\columnwidth}
\includegraphics[width=\textwidth]{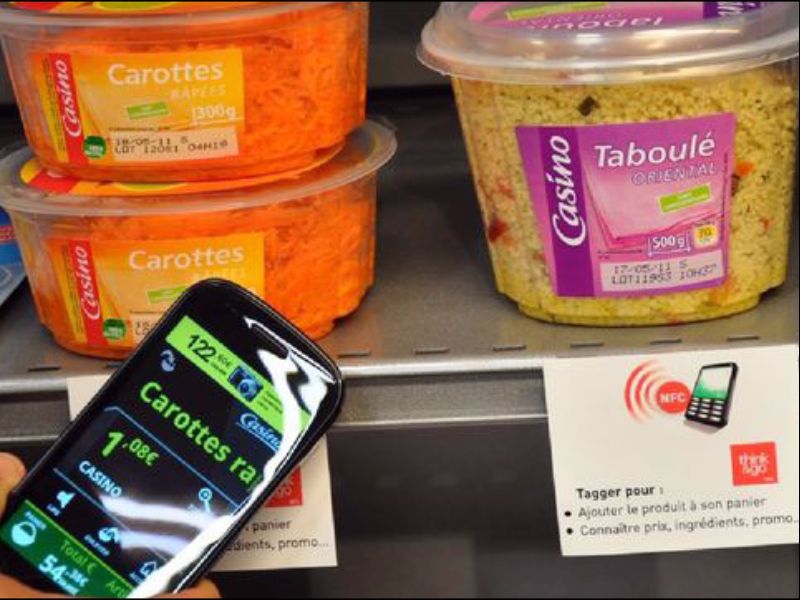}
\caption{Supermarket with NFC tags product labeling (\url{http://www.rfidjournal.com/articles/view?8793/2})}
\label{fig:nfc_supermarket}
\end{subfigure}
\begin{subfigure}[t]{0.5\columnwidth}
\includegraphics[width=\textwidth]{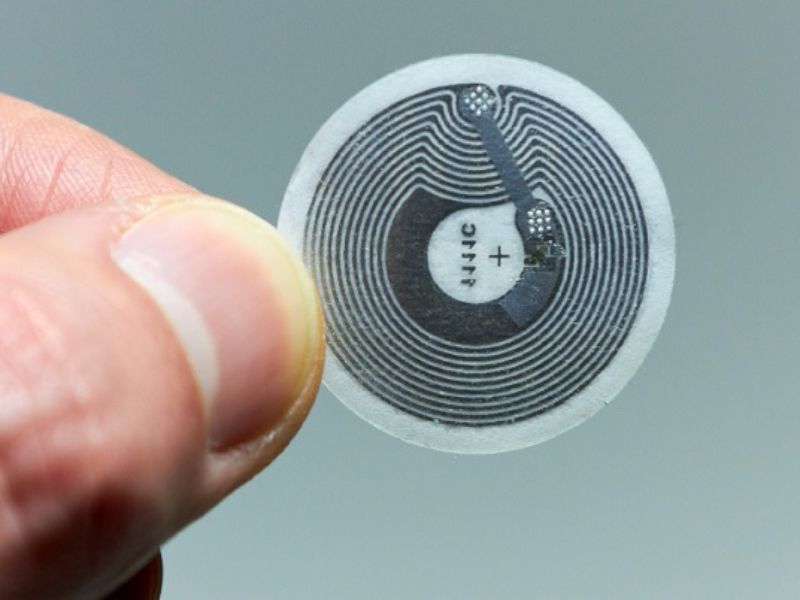}
\caption{NFC-tag closeup (\url{http://www.androidauthority.com/nfc-tags-explained-271872/})}
\label{fig:nfc_tag_closeup}
\end{subfigure}
\caption{NFC application examples}
\label{fig:nfc_applications}
\end{figure*}

Due to the ubiquity of NFC as a fast, simple protocol for small data transactions such as public transport, contactless payments, and building access (hotel rooms) (Figure~\ref{fig:nfc_applications}), we need to be aware of the vulnerabilities that our mobile devices can face. Some of the real world applications require high security measurements to avoid attacks (i.e., payments, location access). In this paper we depict the main vulnerabilities regarding the NFC protocol. Besides we propose some innovative attacks to gather user's device information (fingerprinting), geo-location data gathering without permission requests, social network, bank accounts and Android device hacks. We address these security threads and come up with some solutions to overcome the risks of these contactless communication scenarios.

The rest of the paper is organized as follows: In Section~\ref{sec:nfc_characteristics}, we depict the NFC protocol, modes, and features. Then, in Section~\ref{sec:nfc_vulnerabilities} we describe the NFC vulnerabilities and our findings. In Section~\ref{sec:experiments} we evaluate the found NFC vulnerability and in Section~\ref{sec:discussion} we give some feasible solutions. In Section~\ref{sec:related_work}, we present the literature review and place our research study in the field. Finally, in Section~\ref{sec:conclusion} we conclude the work of our findings and the possible solutions.


\section{NFC Characteristics} \label{sec:nfc_characteristics}

NFC is a set of protocols that enables the wireless communication between two electronic devices, within a distance less than 4 cm. We can see NFC communications in public transport systems, offices buildings as access cards, and also in the commercial contactless payments from VISA, Apple, and Google. 
The NFC protocol consumes very little energy and its transmission speed capabilities are limited to less than 500 Kbps. Although, it is not a protocol to be consider in scenarios with high bandwidth requirements, it has been proposed as another option against QR codes, and channel to create secure WiFi communication between devices (i.e., cameras send the required information to create a secure WiFi connection camera-smartphone). 

There are two type of devices to interact via NFC: \textit{NFC-full devices}, NFC active device that can interact with other NFC peers; \textit{NFC tags}, NFC passive data stores that can be read or written by another NFC-full device. The NFC-full devices can work in three different modes:  \textit{(1) Card emulation mode}: it enables mobile devices such as smartphones to act like an NFC card that an external NFC-reader can access. In example the NFC point-of-sale scenario. \textit{(2) Reader/writer mode}: it enables the NFC device to read/write  NFC-tags. \textit{(3) Peer-to-peer mode}: it allows the NFC device to exchange data with other NFC peers, called Android Beam for Android devices.


NFC support has started since Android\footnote{\url{https://www.android.com/}} version 2.3 (Gingerbread), December, 2010. Android Beam has started in later versions, since 4.0.1 (ICS). More complex NFC modes such as Host Card Emulation (HCE) have been supported since version 4.4.x (KitKat). Android devices need to be unlocked in order to interact with other NFC device\footnote{\url{https://developer.android.com/guide/topics/connectivity/nfc/nfc.html}}. Android Beam allows Android supported devices to transmit data via NFC. Although we will no cover any vulnerability using this Android feature, there exists some risks allowing the transmission of any data just approaching both devices together. The attacker can place an Android device in some non-visible location to the user and activate Android Beam once the user's device are back to back, this action will start the data transmission (i.e., URL, file).

\section{NFC Vulnerabilities} \label{sec:nfc_vulnerabilities}

In this Section we will describe some documented NFC vulnerabilities and other possible protocol weakness of the ecosystem NFC-Android devices. For the latter, due to the myriad of devices and Android OS versions combinations not all the mentioned vulnerabilities affect the devices analogously.

\begin{figure}[t]
\centering
\includegraphics[width =0.8\columnwidth]{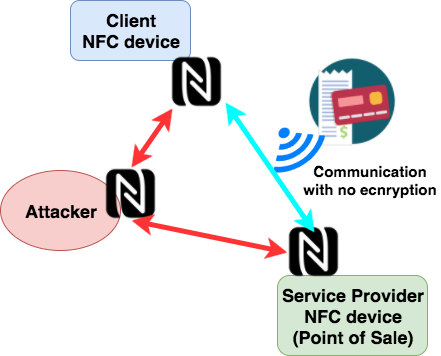}
\caption{NFC Point of sale eavesdropping attack example}
\label{fig:nfc_security_eavesdropping}
\end{figure}

\textit{Lack of security protection of communication}. Most NFC communications do not include encryption mechanisms during its data exchange, it relies on the short range (i.e., less than 4 cm) to guarantee absence of eavesdropping attacks. However, the attacker can still place the device (i.e., NFC tag or NFC reader/writer) between client and NFC provider (i.e., NFC contactless point-of-sale) to trigger a specific attack such as eavesdrop, URL/URI spoofing see Figure~\ref{fig:nfc_security_eavesdropping}. This vulnerability can also be exploited to jam the data exchange between two parties by sending out specific packet at the right timing, which can lead to a deny-of-service (DoS) attack toward the NFC-service provider.

\textit{URL/URI spoofing}. In \cite{mulliner2009:vulnerability}, the authors show that spoofing attacks can be performed to trick the user to see the false information as a valid one. In example, the attacker will design an exactly copy of a user's trusted website with an almost equal URL so the user does not see the difference if she is not cautious. In addition, to uniform resource identifier (URI) and uniform/universal resource locater (URL) spoofing the research shows that phone call and text-message spoofing using the NFC protocol are also applicable. Furthermore, URI and URL spoofing are specially useful in combination with other attacks (i.e., cross-site request forgery).

\begin{figure}[t]
\centering
\includegraphics[width =0.6\columnwidth]{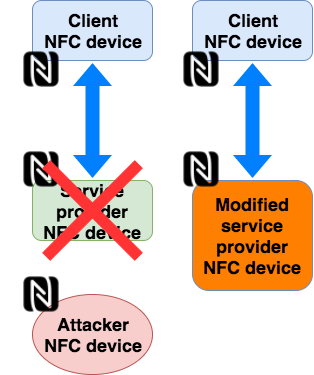}
\caption{NFC tag-device replacement}
\label{fig:nfc_security_tagReplacement}
\end{figure}

\textit{Lack of authentication mechanism of NFC device}. When the NFC reader reads information from another NFC-enabled device, there is not any authentication mechanisms available. Therefore, there is a potential risk of tag replacement and tag hiding (TRTH) attack. In the TRTH scenario, the NFC tags are overwritten by an attacker with malicious information or the physical tag is replace with another tag prepared by the attacker, see Figure~\ref{fig:nfc_security_tagReplacement}.

\textit{Automatic and non-user intervention URL/URI connection}. The proposed attack takes advantage of the non-user intervention when the device detects another NFC device in its proximity. The malicious NFC provides an URL/URI to attack the user's device, as the Android system does not request any user intervention, the device will automatically open the provided link by either other smartphone or NFC-tag. This situation opens security and privacy threads for the device's owner. Once, the device opens the link, it can be attacked by fingerprinting mechanisms or share the user's location for example (see more details in Section~\ref{sec:experiments}. The URI can also open application services such as \textit{contacts} to automatically add malicious contacts without user permission request.

\section{Experiments} \label{sec:experiments}

In this Section we proceed to enumerate the different attacks that can be leveraged using the NFC protocol and Android devices. Due to the myriad of Android devices and the different configurations regarding hardware and software, the proposed and also claimed vulnerabilities in this section can be effective or not. We have tested several device configurations, see Table~\ref{tab:android_devices}.

Android NFC aims to work seamlessly without user intervention. However, once the device is unlocked and NFC enabled, it starts looking for nearby NFC tags in order to read the data stored. This scenario can lead to our proposed attack: open a new channel to attack user's device. The attack can be a fingerprint of the device, gather user's geo-location, or lead to more sophisticated attacks 
such as taking advantage of already logged sites to take actions (i.e., liked a specific web site via Facebook), gather social network information from user's profile. Furthermore, we can use some user's default application permission to access via browsers (i.e., Chrome, Native, Firefox) device's microphone, camera or in cases of documented and not patched security threads: the device's file system. Not every Android user is experienced with technical details, application permissions, and not every Android device is updated with the latest version. Therefore, the scenario that can lead to last mentioned more harmful and privacy concerned (i.e., eavesdropping user surroundings) is feasible. In the following Section we will discuss the situations where the user is aware of the attack (i.e., detects the browsers opening).

The attack can be achieved placing NFC-tags that unlocked Android devices will read in several locations: \textit{(1)} public transport: in areas where the public transport uses NFC reader we can track user's movement from one station to another, collect the user's routine information (i.e., when the user goes to work and back home, where does he work); \textit{(2)} coffee shop, poster at shopping mall : placing NFC-tags under coffee tables or in locations where users tend to leave the device unlocked, we can collect not only device's information but also the geo-location as we know where the tag is located. For both situations, we can also collect the mentioned social network profiles or leverage more complex attacks in combination with other documented browser vulnerabilities.



\begin{table}[t]
\begin{center}
\caption{A table comparing different mobile devices tested}
\begin{tabularx}{\columnwidth}{c c c c}
\toprule
Device   &  Android OS  & Stock Version   & Browsers\\ 
 \cmidrule(r){1-4}
 \multirow{3}{*}{One Plus 3T} & \multirow{3}{*}{Android 7.1.1} & \multirow{3}{*}{Stock} & Chrome\\
 & & & Firefox \\
 & & & Pyro \\
 \multirow{3}{*}{Xiaomi Mi3W} & \multirow{3}{*}{Android 5.1} & \multirow{3}{*}{MIUI 7} & Chrome\\
 & & & Firefox \\
 & & & Native Browser\\
 \multirow{3}{*}{Xiaomi Mi3W} & \multirow{3}{*}{Android 6.0.1} & \multirow{3}{*}{MIUI 8} & Chrome\\
 & & & Firefox \\
 & & & Native Browser\\
 \multirow{3}{*}{Samsung C7} & \multirow{3}{*}{Android 6.0.1} & \multirow{3}{*}{Touchwiz} & Chrome\\
 & & & Firefox \\
 & & & Native Browser\\
 \bottomrule
\end{tabularx}
\label{tab:android_devices}
\end{center}
\end{table}

\begin{lstlisting}[language=bash, basicstyle=\fontsize{6}{11}\ttfamily, label={lst:facebookAPI} , caption={Facebook API like url}]
curl -X POST \
  -F 'access_token=USER_ACCESS_TOKEN' \
  -F 'object=OG_OBJECT_URL' \
https://graph.facebook.com/[User FB ID]/og.likes
\end{lstlisting}

\textit{Facebook Graph API}, the NFC-tag can leverage attacks such as the liking of custom web pages directly from the URL without user intervention (Listing~\ref{lst:facebookAPI}), we only need to send the tag to a device which is already logged in the social network. Other social networks such as Twitter provides web intents to redirect to follow specific web pages (Listing~\ref{lst:twitter_follow}), the attacker can use the \textit{Twitter API} to follow the web pages that she wants.
Some of the current mentioned social APIs, might require user intervention such as in the cases of following a user. There is always the possibility of create a fake account/application to confuse the user and leverage the action with user intervention (permission).

\begin{lstlisting}[language=html, label={lst:twitter_follow} , caption={Twitter follow web intent example}]
https://twitter.com/intent/follow?user_id=2244994945
\end{lstlisting}



There are other possibilities which can lead to more harmful situations than social networks, online banking systems can be the targets of such attacks, with similar methods. The attacker can use the parameters embedded in the URL, and write the URL in the NFC tag. As an example, see Listing~\ref{lst:bank_url}, the attack can perform more sophisticated attacks such as send money from the user account to the attacker one.

\begin{lstlisting}[language=html, label={lst:bank_url} , caption={Bank URL example}]
http://banksite.com/MyAccount/transfer?account="transfer_to"&amount="wanted_amount"
\end{lstlisting}

\begin{figure}[t]
\centering
\begin{subfigure}[t]{0.7\columnwidth}
\includegraphics[width=\textwidth]{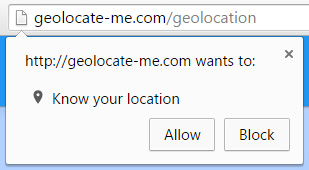}
\caption{Chrome location pop-up}
\label{fig:location_chrome}
\end{subfigure}
\begin{subfigure}[t]{0.7\columnwidth}
\includegraphics[width=\textwidth]{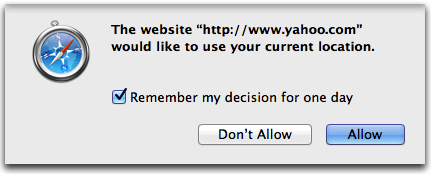}
\caption{Safari location pop-up}
\label{fig:location_safari}
\end{subfigure}
\caption{Browser location-request information pop-ups}
\label{fig:location_popup}
\end{figure}

Another privacy concern is user's geo-location tracking. Modern browsers have already solved this privacy issue with a warning pop-up message to inform the user about the server requesting permission to track user's location, Figure~\ref{fig:location_popup}. The Android-NFC protocol enables an automatic and non-user intervention access to the tagged URL. This issue opens new possibilities to provide user location to services that are interested. The location of the NFC is known, so the attacker only needs to include the geo-location parameters into the URL, once the user's device is near the NFC tag, this will automatically open the user's device browser with the sensitive location information that will be collected by the server. Furthermore, we can include other device's parameters using JavaScript; which is enabled by default in the majority of mobile browsers. As Listing~\ref{lst:javascript_fingerprint} illustrates the URL can track and generate a user fingerprint with the location collected in the URL parameters. In the previous example we have used the library provided by Valve\footnote{https://github.com/Valve/fingerprintjs2}. There is extensive work and real attacks that use the Cookies to track and collect users' information~\cite{kurtz2016:fingerprinting},~\cite{yen2012:host}. URL/URI spoofing can also  add automatically malicious content on user's device such as contacts (text/vcard~\ref{lst:uri_contacts}) without user's intervention.

\begin{lstlisting}[language=html,  basicstyle=\fontsize{5}{11}\ttfamily, label={lst:javascript_fingerprint} , caption={Website that generates a user's fingerprint and collects it in the server side}]
 <!DOCTYPE html>
<html>
  <head>
     <meta charset="UTF-8">
    <title>Example "Fingerprinting" Application</title>
    <script src="https://code.jquery.com/jquery-1.11.3.min.js"></script>
    <script src="https://valve.github.io/fingerprintjs2/fingerprint2.js"></script>
  </head>
  <body>
    <h1>Fingerprint Example</h1>
        <div id="result"></div>
        <code id="components"></code>
        <script>
          console.log("test");
          setTimeout(function () {
              window.location.href = "https://www.google.com"; //will redirect
          }, 200); //will call the function after 200 (X) milliseconds or ready().
          new Fingerprint2().get(function(result, components){
            // this will use all available fingerprinting sources
            console.log(result);
            $('#result').text(JSON.stringify(result));            
            // components is an array of all fingerprinting components used
            console.log(components);
            $('#components').text(JSON.stringify(components));
            //Send components fingerprint to server
            $.ajax({
              url: 'localhost:8882/collectFingerprint',
              type: 'POST',
              contentType:'application/json',
              data: JSON.stringify(result),
              dataType:'json'
            });
          });
        </script>
  </body>
</html>
\end{lstlisting}

\begin{figure}[t]
\centering
\includegraphics[width =0.5\columnwidth]{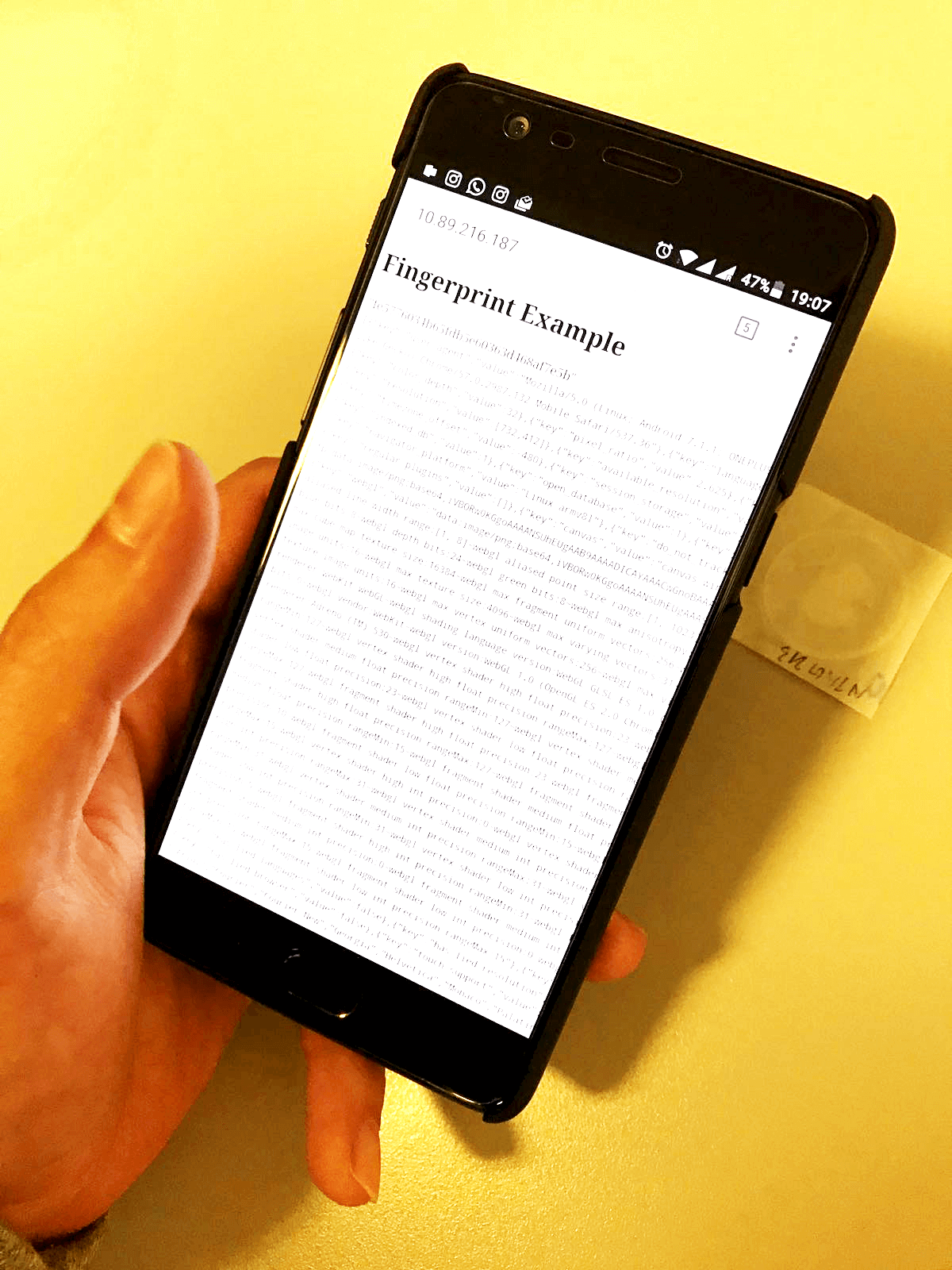}
\caption{Device fingerprinting using the NFC tag}
\label{fig:nfc_fingerprint}
\end{figure}


\begin{lstlisting}[language=bash, basicstyle=\fontsize{5}{11}\ttfamily, label={lst:uri_contacts}, caption={URI VCARD contact, it creates a contact in the user's device without user intervention}]
BEGIN:VCARD
VERSION:4.0
N:MC;Mr.;
FN:Malicious Contact
TEL;TYPE=work,voice;VALUE=uri:tel:+123
EMAIL:maliciouscontact@example.com
END:VCARD
\end{lstlisting}

\begin{lstlisting}[language=bash, basicstyle=\fontsize{5}{11}\ttfamily, label={lst:url_location}, caption={URL track user's location (i.e. latitude and longitude)}]
http://localhost:8888?lat=1&long=3
\end{lstlisting}

\begin{lstlisting}[language=php, basicstyle=\fontsize{5}{11}\ttfamily, label={lst:php_server},caption={PHP server to track device's location and set user's cookie}]
<?php
$value = 'something from somewhere a cookie 3';
setcookie("TestCookie", $value);
$parts = parse_url($url);
echo $_GET["lat"];
echo $_GET["long"];
?>
<html>
 <head>
  <title>PHP Test</title>
 </head>
 <body>
   <?php echo '<p>Hello World</p>'; ?>
   <?php echo $_COOKIE["TestCookie"]; ?>
 </body>
</html>
\end{lstlisting}

\section{Discussion} \label{sec:discussion}

We mentioned some of the vulnerabilities we have found in the Android-NFC ecosystem. In our proposed attack the user can collect user's device information, location, social network profile, or listen to private conversations in situations where the browser permission are enabled. The proposed threads and vulnerabilities can leverage more sophisticated attacks if they are combined with known mobile browser vulnerabilities, or in the case of logged web sites such as Facebook, Banks, the attacker can create defined actions that can take advantage of the already logged situation.

We propose possible solutions to avoid this harmful situations on mobile NFC-enabled environments. One solution to avoid the privacy leakage is in cases where the NFC reads an URL/URI it request user's permission to access the link. Another solution can work analogous as with BLE Beacons, once the device's reads a URL/URI it creates a notification in the notification bar, so the user can access the address when she desires (no work-flow interruption).

\section{Related Work} \label{sec:related_work}

In this literature review we depict the state of art of NFC protocol vulnerabilities, mobile device fingerprinting techniques, and wireless location-based methods using wireless protocols.

The NFC communication protocol is vulnerable to several threads~\cite{chattha2014:nfc}: \textit{eavesdropping}, it is a key thread of wireless communications, as the data transmitted via NFC channel can be intercepted or received by an attacker; \textit{data corruption}, the data transmitted can be modified (corrupted) by an attacker. Denial-of-service (DoS) attacks, and NFC-tag overwriting can be considered within this thread group; \textit{data modification}; \textit{data insertion}, during the exchange of messages data can be inserted; \textit{man-in-the-middle attacks}. Mulliner et al.~\cite{mulliner2009:vulnerability} identify a number of vulnerabilities and threads on NFC-enabled mobile phones: \textit{mobile telephony service attacks}, using URI spoofing the attacker can leverage malicious SMS, telephone receiver number; \textit{URI/URL spoofing}, URL spoofing enabled via NFC communication protocol; \textit{DoS}. In~\cite{chen2014:nfc}, the authors contribute with the enumeration of more possible NFC attacks: \textit{relay attack}, request confidential information from the secure element (more information Section~\ref{sec:nfc_characteristics}); \textit{phising attack}, NFC tag that will execute commands in the user's device(i.e., send email, WiFi AP connection setup); \textit{ticket cloning}, related to the copy of e-tickets. Eun et al.~\cite{eun2013:conditional} propose conditional privacy protections against impersonation attacks from NFC eavesdropping (man-in-the-middle) using the user's public key schemes, pseudonyms and addition trusted third party to protect the privacy of users. 

\begin{figure*}[t]
\centering
\begin{subfigure}[t]{0.6\columnwidth}
\includegraphics[width=\textwidth]{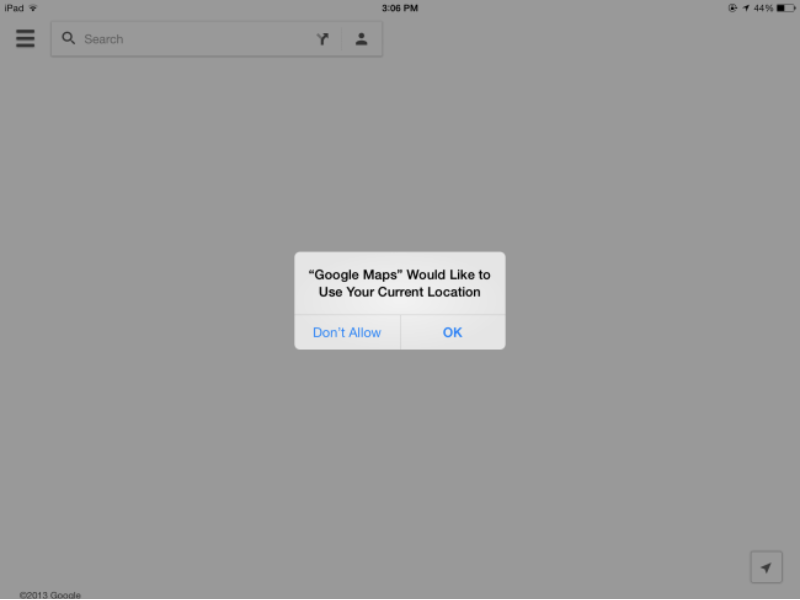}
\caption{iOS app permission requests. The permission requests are shown when the application needs to access the particular restricted resource (after installation) such as contacts, camera, location}
\label{fig:ios_app_permission}
\end{subfigure}
\begin{subfigure}[t]{0.6\columnwidth}
\includegraphics[width=\textwidth]{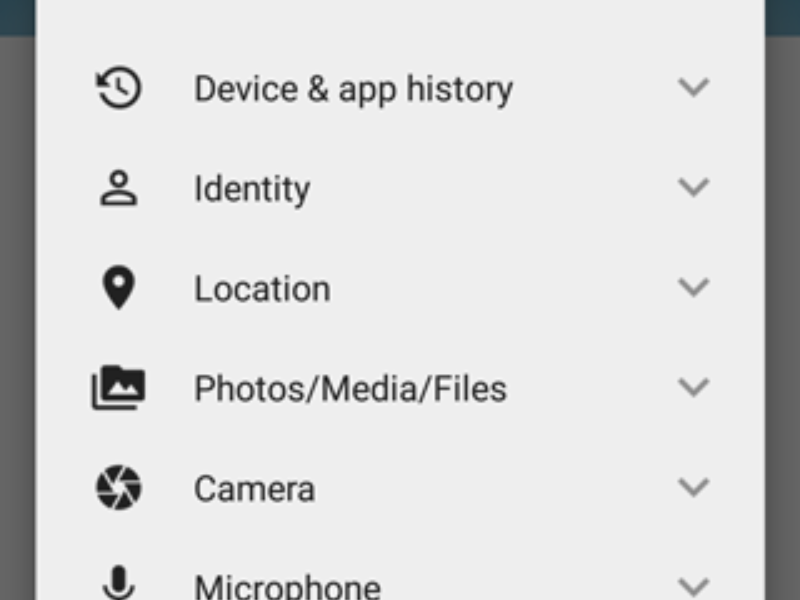}
\caption{Android app permission before Marshmallow (Android 6.0). The app permission are requested before installing the app}
\label{fig:android_install_permission}
\end{subfigure}
\begin{subfigure}[t]{0.6\columnwidth}
\includegraphics[width=\textwidth]{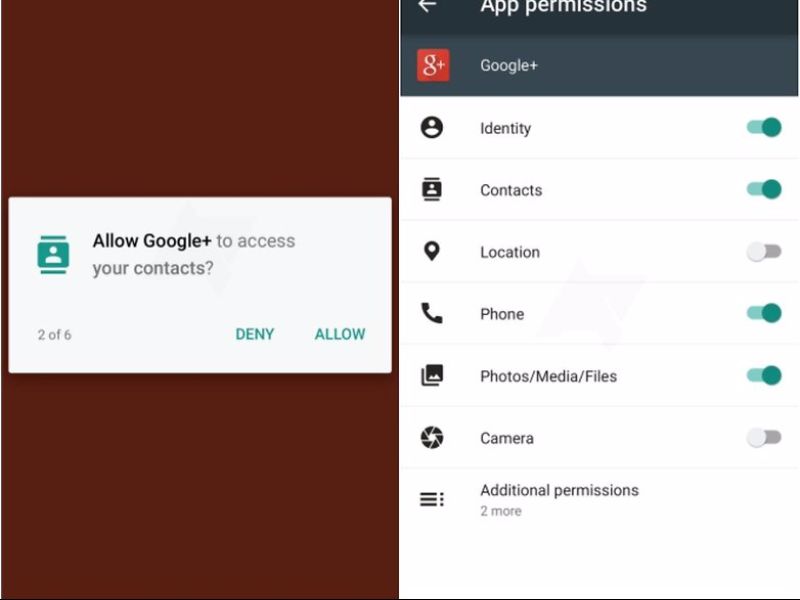}
\caption{Android app permission settings since Marshmallow (Android 6.0). The user can change the app permission in settings menu}
\label{fig:android_m_permission}
\end{subfigure}
\caption{Application permission information}
\label{fig:app_permission}
\end{figure*}

Mobile device's fingerprinting is the information collected via web-based methods (i.e., JavaScript, browser-plugins, cookies), sensor-based (i.e., accelerometer, GPS, WiFi). 
Hupperich et al.~\cite{hupperich2015:robustness} propose a system with modern web-based fingerprinting techniques, for mobile devices. The authors discover that some features used in desktop environments lose their importance in mobile fingerprinting environments. They also test the proposed system against evasion attacks such as changeability of features (i.e., use of a second browser, proxy). 
The authors mention that the browser plugins and extensions, due to the complexity of commercial browsers, are the biggest thread in device's web-based fingerprinting. AppPrint learns mobile app fingerprints from traffic observations~\cite{miskovic2015:appprint}. The system is capable of identifying with high granularity installed mobile applications based on the network traffic they generate. However, the deployment of the system is situated in the network provider side (i.e., network administrator). In~\cite{bojinov2014:mobile}, the authors present a sensor-based fingerprinting approach using speaker and accelerometer's device. The authors' work raises open problems regarding the fingerprinting using other mobile device's sensors. 
Kurtz et al. \cite{kurtz2016:fingerprinting} present an innovative software-based approach to fingerprinting mobile devices using the user's personalized configurations. The proposed approach uses a third party application, instead of the most used web browser channel, to obtain a unique device fingerprint. 
However, in third party application approaches for Android and iOS platforms the user needs to grant app permission in order for the app to access the sensors (see Figure~\ref{fig:app_permission}). 

Furthermore, fingerprinting of user's location is a very important feature, that we have described with the sensor-based fingerpringting related work. Although, there are other approaches to track user's indoor location as we will enumerate. There are myriad related work on 
location-based methods using wireless such as WiFi, Bluetooth, ultra wide band (UWB) sensors (\cite{yang2015:wifi},~\cite{ruiz2017:comparing}).
In this paper~\cite{bisio2016:smart}, the authors present a computational approach to speedup tradition indoor WiFi location-based methods. 
Although these methods can track device's location within a particular environment (i.e., shopping malls) using wireless methods, it lacks of user's traceability and it requires installation of the system either on the user's device, or access point (AP) mesh networks.


\section{Conclusion} \label{sec:conclusion}

In this paper we depict the current state of mobile NFC-enabled ecosystem's vulnerabilities and threads. Some previous work attacks still can be enabled in current NFC communications and the situation of application permission on Android devices. The latter is still a difficult scenario due to the differences between OS versions, how the OS approaches the permission requests (before app installation), and the non technical experience Android users. Furthermore, the location-based tracking attempts have been solved by browser location-request notifications. However, NFC-tags provide a simple and non-user intervention channel to track user's location, fingerprinting, and other logged-based web sites attacks. In summary, our proposed attack enables current mobile fingerprinting techniques using the NFC-tags and URL/URI Android mobile device's vulnerabilities. To conclude, we propose simple deployable solutions that will not interrupt the user-workflow to enable a secure and privacy aware NFC-Android ecosystem.



\bibliographystyle{IEEEtran}
\bibliography{nfc_security}

\vfill

\eject

\IEEEoverridecommandlockouts

\begin{IEEEbiography}[{\includegraphics[width=1in,height=1.25in,clip,keepaspectratio]{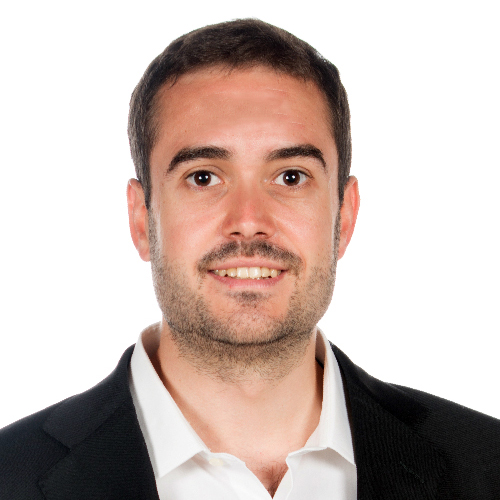}}]{Carlos Bermejo}
received his MSc. degree in Telecommunication Engineering in 2012 from Oviedo university, Spain. He is currently a PhD student at Hong Kong University of Science and Technology working at the Symlab research group. His main research interest are Internet-of-Things, mobile augmented reality, network security, human-computer-interaction, social networks, and device-to-device communication.
\end{IEEEbiography}

\vspace{-1cm}

\begin{IEEEbiography}[{\includegraphics[width=1in,height=1.25in,clip,keepaspectratio]{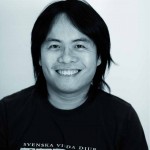}}]{Pan Hui}
received his Ph.D degree from Computer Laboratory, University of Cambridge, and earned his MPhil and BEng both from the Department of Electrical and Electronic Engineering, University of Hong Kong. He is currently a faculty member of the Department of Computer Science and Engineering at the Hong Kong University of Science and Technology where he directs the HKUST-DT System and Media Lab ( http://symlab.ust.hk ). He is an adjunct Professor of social computing and networking at Aalto University Finland and also served as a Distinguished Scientist for Telekom Innovation Laboratories (T-labs) Germany until 2015. Before returning to Hong Kong in 2013, he has spent several years in T-labs and Intel Research Cambridge. He has published over 200 research papers with over 12,000 citations and has around 30 granted / filed European patents. He has founded and chaired several IEEE/ACM conferences/workshops, and has been serving on the organising and technical program committee of numerous international conferences including ACM SIGCOMM, IEEE Infocom, ICNP, SECON, MASS, Globecom, WCNC, ITC, IJCAI, ICWSM and WWW. He is an associate editor for IEEE Transactions on Mobile Computing and IEEE Transactions on Cloud Computing, and an ACM Distinguished Scientist.
\end{IEEEbiography}

\end{document}